\documentstyle[11pt]{article}
\def\fnote#1#2{\begingroup\def\thefootnote{#1}\footnote{#2}\endgroup}

\begin{document}
\hfill{UTTG-07-00}
\baselineskip=36pt
\Large
\begin{center}

The Cosmological Constant Problems\fnote{*}{This research was supported in part by the Robert A. Welch Foundation and NSF Grant PHY-9511632.}\\
\normalsize 
(Talk given at Dark Matter 2000, Marina del Rey, CA, February 2000)\\
\vspace*{15pt}
Steven Weinberg\\
Department of Physics, University of Texas\\ Austin, Texas 78712
\end{center}

\begin{abstract}
The old cosmological constant problem is to understand why the vacuum energy is 
so small; the new problem is to understand why it is comparable to the present 
mass density.  Several approaches to these problems are reviewed.  Quintessence 
does not help with either; anthropic considerations offer a possibility of 
solving both.
In theories with a scalar field that takes random initial values,
the anthropic principle may apply to the cosmological constant, but probably to 
nothing else.
\end{abstract}

\normalsize 
\noindent
1. {\bf Introduction}

There are now two cosmological constant problems.  The old cosmological constant 
problem is to understand in a natural way why the vacuum energy density $\rho_V$ 
is not very much larger.  We can reliably calculate some
contributions to $\rho_V$, like the energy density in fluctuations in the 
gravitational field at graviton energies nearly up to the Planck scale, which is 
larger than is observationally allowed by some 120 orders of magnitude.  Such 
terms in $\rho_V$ can be cancelled by other contributions that we can't 
calculate, but the cancellation then has to be accurate to 120 decimal places.  
The new cosmological constant problem is to understand why $\rho_V$ is not only 
small, but also, as current Type Ia supernova observations seem to 
indicate,\footnote{ A. G. Riess {\em et al.}: Astron. J. {\bf 116}, 1009 (1998): P. M. 
Garnavich {\em et al.}: Astrophys. J. {\bf 509}, 74 (1998); S. Perlmutter {\em et al.}: Astrophys. J. {\bf 517}, 565 
(1999).} of the same order 
of magnitude as the present mass density of the universe.  

The efforts to understand these problems can be grouped into four general 
classes.
The first approach is to imagine some scalar field coupled to gravity in such a
way that $\rho_V$ is automatically cancelled or nearly cancelled when the scalar 
field reaches its equilibrium value.  In a review article over a decade 
ago\footnote{S. Weinberg: Rev. Mod. Phys. {\bf 61}, 1 (1989).} I gave a sort of 
`no go' theorem, showing why such attempts would not work without the need for a 
fine tuning of parameters that is just as mysterious as the problem we started 
with.   I wouldn't claim that this is conclusive ---
other no-go theorems have  been evaded in the past --- but so far no one has 
found a way out of this one.  The second approach is to imagine some sort of 
deep symmetry, one that is not apparent in the effective field theory that 
governs phenomena at accessible energies, but that nevertheless constrains the 
parameters of this effective theory so that $\rho_V$ is zero or very small.  I 
leave this to be covered in the talk by Edward Witten.  In this talk I will 
concentrate on the third and fourth of these approaches, based respectively on 
the idea of quintessence and on versions of the anthropic principle.

\vspace{20pt}
\noindent
2. {\bf Quintessence}

The idea of quintessence\footnote{P. J. E. Peebles and B. Ratra: Astrophys. J. 
{\bf 325}, L17 (1988); B. Ratra and P. J. E. Peebles: Phys. Rev. {\bf D 37}, 
3406 (1988); C. Wetterich: Nucl. Phys. {\bf B302}, 668 (1988).} is that the 
cosmological constant is small because the universe is old.  One imagines a 
uniform scalar field $\phi(t)$ that rolls down
a potential $V(\phi)$, at a rate governed by the field equation 
\begin{equation}
\ddot{\phi}+3H\dot{\phi}+V'(\phi)=0\;,
\end{equation}
where $H$ is the expansion rate
\begin{equation}
H=\sqrt{\left(\frac{3}{8 \pi G}\right)\left(\rho_\phi+\rho_M\right)}\;.
\end{equation}
Here $\rho_\phi$ is the energy density of the scalar field
\begin{equation} 
\rho_\phi=\frac{1}{2}\dot{\phi}^2+V(\phi)\;,
\end{equation} 
while $\rho_M$ 
is the energy density of matter and radiation,  which decreases as
\begin{equation} 
\dot{\rho}_M=-3H\,\left(\rho_M+p_M\right)\;,
\end{equation} 
with $p_M$ the pressure of matter and radiation.

 If there is some 
value of $\phi$ (typically, $\phi$ infinite) where $V'(\phi)=0$, then it is 
natural that $\phi$ should approach this value, so that it eventually changes 
only slowly with time.  Meanwhile $\rho_M$ is steadily decreasing, so that 
eventually the universe starts an exponential expansion with a slowly varying 
expansion rate $H\simeq\sqrt{8\pi G V(\phi)/3}$. The problem, of course, is to 
explain why $V(\phi)$ is small or zero at the value of $\phi$ where 
$V'(\phi)=0$.  

Recently this approach has been studied in the context of so-called `tracker' 
solutions.\footnote{I. Zlatev, L. Wang, and P. J. Steinhardt: Phys. Rev. Lett. 
{\bf 82}, 896 (1999); Phys. Rev. {\bf D 59}, 123504 (1999).}  The simplest case 
arises for a potential of the form
\begin{equation} 
V(\phi)=M^{4+\alpha}\phi^{-\alpha}\;,
\end{equation} 
where $\alpha>0$, and $M$ is an adjustable constant.  If the scalar field begins 
at a value much less than the Planck mass and with $V(\phi)$ and $\dot{\phi}^2$ 
much less than $\rho_M$, then the field $\phi(t)$ initially increases as 
$t^{2/(2+\alpha)}$, so that $\rho_\phi$ decreases as $t^{-2\alpha/(2+\alpha)}$, 
while $\rho_M$ is  decreasing faster, as $t^{-2}$.  (The existence of this phase 
is important, because the success of cosmic nucleosynthesis calculations would 
be lost if the cosmic energy density were not dominated by $\rho_M$ at 
temperatures of order $10^9\;{}^\circ$K to $10^{10}\;{}^\circ$K.)   Eventually a 
time is reached when $\rho_M$ becomes as small as $\rho_\phi$, after which the 
character of the solution changes.  Now  $\rho_\phi$ becomes larger than 
$\rho_M$, and $\rho_\phi$ decreases more slowly, as $t^{-2/(4+\alpha)}$.  The 
expansion rate $H$ now goes
as $H\propto \sqrt{V(\phi)} \propto t^{-\alpha/(4+\alpha)}$, so the 
Robertson--Walker scale factor $R(t)$ grows almost exponentially, with $\log 
R(t)\propto t^{4/(4+\alpha)}$.  In this approach, the transition from 
$\rho_M$-dominance to $\rho_\phi$-dominance is supposed to take place near the 
present time, so that both $\rho_M$ and $\rho_\phi$ are now both contributing 
appreciably to the cosmic expansion rate.

The nice thing about these tracker solutions is that the
existence of a cross-over from an early  $\rho_M$-dominated expansion to a later 
$\rho_\phi$-dominated expansion does not depend on any fine-tuning of the 
initial conditions.  But it should not be thought that {\em either} of the two 
cosmological constant problems are solved in this way.  Obviously, the decrease 
of $\rho_\phi$ at late times would be spoiled if we added a constant of order 
$m_{\rm Planck}^4$ (or $m_W^4$, or $m_e^4$) to the potential (5).  What is 
perhaps less clear is that, even if we take the potential in the form (5) 
without any such added constant, we still need a fine-tuning to make the value 
of $\rho_\phi$ at which $\rho_\phi\approx \rho_M$ close to the {\em present} 
critical density $\rho_{c0}$.   The value of the field $\phi(t)$ at this 
crossover can easily be seen to be of the order of the Planck mass, so in order 
for $\rho_\phi$ to be comparable to $\rho_M$ at the present time we need
\begin{equation} 
M^{4+\alpha}\approx (8\pi G)^{-\alpha/2}\rho_{c0}\approx (8\pi G)^{-1-
\alpha/2}H_0^2\;.
\end{equation} 
Theories of quintessence offer no explanation why this should be the case.
(An interesting suggestion has been made after Dark Matter 2000.\footnote{ C. Armendariz-Picon, V. Mukhanov, and P. J. Steinhardt: 
astro-ph/0004134.})

\vspace{20pt}
\noindent
3. {\bf Anthropic Considerations}

In several  cosmological theories the observed big bang is just one member of an 
ensemble.  The ensemble may consist of different expanding regions at different 
times and locations in the same spacetime,\footnote{A. Vilenkin: Phys. Rev. {\bf 
D 27}, 2848 (1983); A. D. Linde: Phys. Lett. {\bf B175}, 395 (1986).} or of 
different terms in the wave function of the universe.\footnote{E. Baum: Phys. 
Lett. {\bf B133}, 185 (1984); S. W. Hawking: in {\em Shelter Island II -- 
Proceedings of the 1983 Shelter Island Conference on Quantum Field Theory and 
the Fundamental Problems of Physics}, ed. by R. Jackiw {\em et al.} (MIT Press, 
Cambridge, 1985); Phys. Lett. {\bf B134}, 403 (1984); S. Coleman: Nucl. Phys. 
{\bf B 307}, 867 (1988).}  If the vacuum energy density $\rho_V$ varies among 
the different members of this ensemble, then the value observed by any species 
of astronomers  will be conditioned by the necessity that this value of $\rho_V$ 
should be suitable for the evolution of intelligent life.  

It would be a disappointment if this were the solution of the cosmological 
constant problems, because we would like to be able to calculate all the 
constants of nature from first principles, but it may be a disappointment that 
we will have to live with.  We have learned to live with similar disappointments 
in the past.  For instance, Kepler tried to derive the relative distances of the 
planets from the sun by a geometrical construction involving Platonic solids 
nested within each other, and it was somewhat disappointing when Newton's theory 
of the solar system failed to constrain the radii of planetary orbits, but by 
now we have gotten used to the fact that these radii are what they are because 
of historical accidents.  This is a pretty good analogy, because we do have an 
anthropic explanation why the planet on which we live is in the narrow range of 
distances from the sun at which the  surface temperature allows the existence of 
liquid water:  if the radius of our planet's orbit was not in this range, then 
we would not be here.  This would not be a satisfying explanation if the earth 
were the only planet in the universe, for then the fact that it is just the 
right distance from the sun to allow water to be liquid on its surface would be 
quite amazing.  But with nine planets in our solar system and vast numbers of 
planets in the rest of the universe, at different distances from their 
respective stars, this sort of anthropic explanation is just common sense.  In 
the same way, an anthropic explanation of the value of $\rho_V$ makes sense if 
and only if there is a very large number of big bangs, with different values for 
$\rho_V$. 

The anthropic bound on a positive vacuum energy density is set by the 
requirement that $\rho_V$ should not be so large as to prevent the formation of 
galaxies.\footnote{S. Weinberg: Phys. Rev. Lett. {\bf 59}, 2607 (1987).}  Using 
the simple spherical infall model of Peebles\footnote{P. J. E. Peebles: 
Astrophys. J. {\bf 147}, 859 (1967).} to follow the nonlinear growth of 
inhomogeneities in the matter density, one finds an upper bound
\begin{equation}
\rho_V<\frac{500\,\rho_R\,\delta_R^3}{729}
\end{equation}
where $\rho_R$ is the mass density and $\delta_R$ is a typical fractional 
density perturbation, both taken  at the time of recombination.  
This is roughly the same as requiring that $\rho_V$ should be no larger than the 
cosmic mass density at the earliest time of galaxy formation, which for a 
maximum galactic redshift of 5 would be about 200 times the present mass 
density.  This is a big improvement over missing by 120 orders of magnitude, but 
not good enough.

However, we would not expect to live in a big bang in which galaxy formation is 
just barely possible.  Much more reasonable is what Vilenkin calls a {\em 
principle of mediocrity},\footnote{A. Vilenkin: Phys. Rev. Lett. {\bf 74}, 846 
(1995); in {\em Cosmological Constant and the Evolution of the Universe}, ed. by 
K. Sato {\em et al.} (Universal Academy Press, Tokyo, 1996).}  which suggests 
that we should expect to find ourselves in a big bang that is typical of those 
in which intelligent life is possible.  To be specific, if ${\cal P}_{\rm 
a\;priori}(\rho_V)\,d\rho_V$ is the {\em a priori} probability of a particular 
big bang having vacuum energy density between $\rho_V$ and $\rho_V+d\rho_V $, 
and ${\cal N}(\rho_V)$ is
the average number of scientific civilizations in big bangs with energy density
$\rho_V$, then the actual (unnormalized) probability of a scientific 
civilization observing an energy density between $\rho_V$ and $\rho_V+d\rho_V $ 
is
\begin{equation} 
d{\cal P}(\rho_V)= {\cal N}(\rho_V)\,{\cal P}_{\rm 
a\;priori}(\rho_V)\,d\rho_V\;.
\end{equation}
We don't know  how to calculate ${\cal N}(\rho_V)$, but it seems reasonable to 
take it as proportional to the number of baryons that wind up in galaxies, with 
an unknown proportionality factor that is independent of $\rho_V$.
There is a complication, that the total number of baryons in a big bang may be 
infinite, and may also depend on $\rho_V$.  In practice, we take ${\cal 
N}(\rho_V)$ as the {\em fraction} of baryons that wind up in galaxies, which we 
can hope to calculate, and include the total baryon number as a factor in ${\cal 
P}_{\rm a\;priori}(\rho_V)$.

The one thing that offers some hope of actually calculating $ d{\cal P}(\rho_V)$ 
is that ${\cal N}(\rho_V)$ is non-zero in only a narrow range of values of 
$\rho_V$, values that are much smaller than the energy densities typical of 
elementary particle physics, so that ${\cal P}_{\rm a\;priori}(\rho_V)$ is 
likely to be constant within this range. \footnote{S. 
Weinberg: in {\em Critical Dialogs in Cosmology}, ed. by N. Turok (World 
Scientific, Singapore, 1997).}  The value of this constant is fixed by 
the requirement that the total probability should be one, so
\begin{equation} 
d{\cal P}(\rho_V)= \frac{{\cal N}(\rho_V) \,d\rho_V}{\int {\cal N}(\rho'_V) 
\,d\rho'_V}\;.
\end{equation}
The fraction ${\cal N}(\rho_V) $ of baryons in galaxies has been calculated by 
Martel, Shapiro and myself,\footnote{H. Martel, P. Shapiro, and S. Weinberg:
Astrophys. J. {\bf 492}, 29 (1998).} using the well-known spherical infall model 
of Gunn and Gott,\footnote{J. Gunn and J. Gott: Astrophys. J. {\bf 176}, 1 
(1972).}  in which one starts with a fractional density perturbation that is 
positive within a sphere, and compensated by a negative fractional density 
perturbation in a surrounding spherical shell.  The results are quite 
insensitive to the relative radii of the sphere and shell.  Taking the shell 
thickness to equal the sphere's radius, the integrated probability distribution 
function for finding a vacuum energy less than or equal to $\rho_V$ is
\begin{eqnarray}
{\cal P}(\leq \rho_V)&\equiv & \int_0^{\rho_V}d{\cal P}\nonumber\\
&=& 1+(1+\beta)e^{-\beta}\nonumber\\
&+&\frac{1}{2\ln 2-1}\int_\beta^\infty e^{-x}dx\left\{-2\sqrt{\beta x}
+\beta+2x\ln\left[\sqrt{\beta/x}+1\right]\right\}~~~~
\end{eqnarray}
where 
\begin{equation}
\beta\equiv 
\frac{1}{2\sigma^2}\left(\frac{729\;\rho_V}{500\;\rho_R}\right)^{2/3}\;~~~~
\end{equation} 
with $\sigma$ the rms fractional density perturbation at recombination, 
and $\rho_R$ the average mass density at recombination.   The probability of 
finding ourselves in a big bang with a vacuum energy density large enough to 
give a present value of $\Omega_V$ of 0.7 or less turns out to be 5\% to 12\%, 
depending on the assumptions used to estimate $\sigma$.  In other words, the 
vacuum energy in our big bang still seems a little low, but not implausibly so.
{\em These anthropic considerations can therefore provide a solution to both the 
old and the new cosmological constant problems,} provided of course that the 
underlying assumptions are valid.
Related anthropic calculations have been carried out by several other 
authors.\footnote{G. Efstathiou: Mon. Not. Roy. Astron. Soc. {\bf 274}, L73 
(1995); M. Tegmark and M. J. Rees: Astrophys. J. {\bf 499}, 526 (1998); J. 
Garriga, M. Livio, and A. Vilenkin: astro-ph/9906210; S. Bludman: 
astro-ph/0002204.}

I should add that when anthropic considerations were first applied to the 
cosmological constant, counts of galaxies as a function of redshift\footnote{E. 
D. Loh: Phys. Rev. Lett. {\bf 57}, 2865 (1986).} indicated that $\Omega_\Lambda$ 
is $0.1^{+0.2}_{-0.4}$, and this was recognized to be too small to be explained 
anthropically.  The subsequent discovery in studies of type Ia supernova 
distances and redshifts that $\Omega_\Lambda$ is quite large does not of course 
prove that anthropic considerations are relevant, but it is encouraging.

Recently the assumptions underlying these calculations have been challenged by 
Garriga and Vilenkin.\footnote{J. Garriga and A. Vilenkin: astro-ph/9908115.}  
They adopt a plausible model for generating an ensemble of big bangs with 
different values of $\rho_V$, by supposing that there is a scalar field $\phi$ 
that initially can take values anywhere in a broad range in which the potential 
$V(\phi)$ is very flat.  Specifically, in this range
\begin{equation} 
\left|\frac{V'(\phi)}{V(\phi)}\right|\ll\sqrt{8\pi G}~~~~{\rm and}~~~~ 
\left|\frac{V''(\phi)}{V(\phi)}\right|\ll 8\pi G\;.
\end{equation} 
It is also assumed that in this range $V(\phi)$ is much less than the initial 
value of the energy density $\rho_M$ of matter and radiation.     
For initial values of $\phi$ in this range, the vacuum energy density 
$\rho_\phi$ stays roughly constant while $\rho_M$ drops to a value of order 
$\rho_\phi$.
To see this, note that during this period the expansion rate behaved as 
$H=\eta/t$, with $\eta=2/3$ or $\eta=1/2$ during times of matter or radiation 
dominance, respectively.  If we tentatively assume that $\phi$ is roughly 
constant, then the field equation (1) gives 
\begin{equation}
\dot{\phi}\simeq -\frac{t\,V'(\phi)}{1+3\eta}\;.
\end{equation}
During the time that $\rho_M\gg \rho_\phi$, the ratio of the kinetic to the 
potential terms in Eq.~(3) for $\rho_\phi$ is
\begin{equation}
\frac{\dot{\phi}^2}{2V(\phi)} \simeq \frac{t^2V'^2(\phi)}{2(1+3\eta)^2 V(\phi)}
\ll \frac{8\pi G t^2 \,V(\phi)}{ 2(1+3\eta)^2}\simeq \frac{3\eta^2 \,V(\phi)}{ 
2(1+3\eta)^2\rho_M}\ll 1\;,
\end{equation} 
so $\rho_\phi$ is dominated by the potential term.  The fractional change in 
$\rho_\phi$ until the time $t_c$ when $\rho_M$ becomes equal to $\rho_\phi$ is 
then
\begin{equation}
\frac{|\Delta \rho_\phi |}{\rho_\phi}=\frac{1}{\rho_\phi}\left|\int_0^{t_c} 
V'(\phi)\,\dot{\phi}\,dt\right|\simeq 
\frac{V'^2(\phi)t_c^2}{2(1+3\eta)\rho_\phi}
\approx \frac{V'^2(\phi) }{8\pi G\rho^2_\phi}\ll 1\;.
\end{equation} 
Following this period, $\rho_\phi$ becomes dominant, and the inequalities (12) 
ensure that the expansion becomes essentially exponential, just as in theories 
with the `tracker' solutions discussed in the previous section.  Hence in this 
class of models, $V(\phi)$ plays the role of  a constant vacuum energy, whose 
values are governed by the {\em a priori}
probability distribution for the initial values of $\phi$.  In particular, if 
one assumes that all initial values of $\phi$ are equally probable, then the 
{\em a priori} distribution of the vacuum energy is
\begin{equation}
{\cal P}_{\rm a\;priori}\Big(V(\phi)\Big)\propto 
\frac{1}{\left|V'(\phi)\right|}\;.
\end{equation}

The point made by Garriga  and Vilenkin was that, because $V(\phi)$ is so flat, the field 
$\phi$ can vary appreciably even when $\rho_V\simeq V(\phi)$ is restricted to 
the very narrow anthropically allowed range of values in which galaxy formation 
is possible.   They concluded that it would also be possible for the {\em a 
priori}
probability (16) to vary appreciably in this range, which if true would require 
modifications in the calculation of ${\cal P}(\leq \rho_V)$ described above.  
The potential they used as an example was 
$$ V(\phi)=V_1+A(\phi/M)+B\sin\Big(\phi/M\Big) \;,$$
with $V_1$ large, of order $M^4$, $A$ and $B$ much smaller, and $M$ a large 
mass, but not larger than the Planck mass.  This yields an {\em a priori} 
probability distribution (16) that varies appreciably in the anthropically 
allowed range of $\phi$.

It turns out\footnote{S. Weinberg: astro-ph/0002387.} that the issue of whether 
the {\em a priori}
probability (16) is flat in the anthropically allowed range of $\phi$ depends on 
the way we impose the slow roll conditions (12).  There is a large class of 
potentials for which the probability is flat in this range.  Suppose for 
instance  that, unlike the example chosen by Garriga and Vilenkin, the potential 
is of the general form 
\begin{equation} 
V(\phi)=V_1f(\lambda\phi)
\end{equation}
where $V_1$ is a large energy density, in the range $m_W^4$ to $m_{\rm 
Planck}^4$,  $\lambda>0$ is a very small constant, and $f(x)$ is a function 
involving no very small or very large parameters. Anthropically allowed  values 
of $\phi/\lambda$ must be near a zero of $f(x)$, say a simple zero at $x=a$.  
Then
$V'(\phi)\simeq \lambda V_1 f'(a)\approx \lambda V_1$ and $V''(\phi)\simeq 
\lambda^2 V_1 f''(a)\approx \lambda^2 V_1$, so both inequalities (12) are 
satisfied if 
\begin{equation} 
\lambda\ll \sqrt{8\pi G}\left(\frac{\rho_V}{V_1}\right)\;.
\end{equation}
Galaxy formation is only possible for $|V(\phi)|$ less than an upper bound 
$V_{\rm max}$, of the order of the mass density of the universe at the earliest 
time of galaxy formation, which is very much less than $V_1$, so the 
anthropically allowed range of values of $\phi$ is
\begin{equation}
\left|\phi-a/\lambda\right|_{\rm max}\simeq\frac{V_{\rm max}}{\lambda 
V_1|f'(a)|}\;.
\end{equation} 
The fractional variation in the {\em a priori} probability density (16)
as $\phi$ varies in the range (19) is then
\begin{equation}
\left|\frac{V''(\phi)}{V'(\phi)}\right|\left|\phi-a/\lambda\right|_{\rm max}
\simeq \left|\frac{V_{\rm 
max}}{V_1}\right|\left|\frac{f''(a)}{f'^2(a)}\right|\approx \left|\frac{V_{\rm 
max}}{V_1}\right|\ll 1
\end{equation}
justifying the assumptions made in the calculation of Eq.~(10).

I should emphasize that no fine-tuning is needed in potentials of type (16).  It 
is only necessary that $V_1$ be sufficiently large, $\lambda$ be sufficiently 
small, and $f(x)$ have a simple zero somewhere, with derivatives of order unity 
at this zero.  These properties are not upset if for instance we add a large 
constant to the potential.  But why should each appearance of the field $\phi$ 
be accompanied with a tiny factor $\lambda$?  As we have been using it, 
derivatives of the field $\phi$ appear in the Lagrangian density in the form $-
\frac{1}{2}\partial_\mu\phi
\partial^\mu\phi$, as shown by the coefficient unity of the second derivative in 
the field equation (1).  In general, we might expect the Lagrangian density for
$\phi$ to take the form
\begin{equation}
{\cal L}=-\frac{Z}{2}\partial_\mu\phi\partial^\mu\phi-V_1f(\phi/M)
\end{equation}
where $f(x)$ is a function of the sort we have been considering, involving no 
large or small parameters, $M$ is a mass perhaps of order $(8\pi G)^{-1/2}$, and 
$V_1$ is a large constant, of order $M^4$.  With an arbitrary 
field-renormalization constant $Z$ in the Lagrangian, the field $\phi$ is not 
canonically normalized, and does not obey Eq.~(1).  We may define a canonically 
normalized field as $\phi'\equiv \sqrt{Z}\phi$; writing the Lagrangian in terms 
of $\phi'$, and dropping the prime, we get a potential of the form (16), with 
$\lambda=1/M\sqrt{Z}$.  Thus we can understand a very small $\lambda$ if we can 
explain why the field renormalization constant $Z$ is very large.  Perhaps this 
has something to do with the running of $Z$ as the length scale at which it is 
measured grows to astronomical dimensions.

There is a problem with this sort of implementation of the anthropic principle, 
that may prevent its application to anything other than the cosmological 
constant.  When quantized, a scalar field with a very flat potential leads to 
very light bosons, that might be expected to have been already observed.  If we 
want to explain the masses and charges of elementary particles anthropically, by 
supposing that these masses and charges arise from expectation values of a 
scalar field in a flat potential with random initial values, then the scalar 
field would have to couple to these elementary particles, and would therefore be 
created in their collisions and decays.  This problem does not arise for a 
scalar field that couples only to itself and gravitation (and perhaps also to a 
hidden sector of other fields that couple only to other fields in the hidden 
sector and to gravitation).  It is true that
such a scalar would couple to observed particles through multi-graviton 
exchange, and with a cutoff at the Planck mass the Yukawa couplings of 
dimensionality four that are generated in this way would in general not be 
suppressed by factors of $G$.  But in our case the 
non-derivative interactions of the scalars with gravitation are suppressed by
a factor $V'(\phi)\propto\lambda$, which according to Eq.~(18) is much less than
$\sqrt{8\pi G}$, yielding Yukawa couplings that are very much less than unity.
Thus it may be that anthropic considerations are relevant for the cosmological 
constant, but for nothing else.

\end{document}